\documentclass[a4paper,12pt]{article}
\usepackage[hmargin=2.5cm,vmargin=2.5cm]{geometry}
\usepackage[colorlinks,urlcolor=blue,citecolor=blue]{hyperref}
\usepackage{subcaption}
\usepackage{graphicx}

\usepackage{authblk}
\usepackage{bbm}
\usepackage{physics}

\title{Matrix product solution of the stationary states of two-species open zero range processes}
\author[1]{Zhongtao Mei\thanks{zhongtao.mei@apctp.org}} 
\author[1,2]{Jaeyoon Cho}
\affil[1]{Asia Pacific Center for Theoretical Physics, Pohang 37673, Korea}
\affil[2]{Department of Physics, POSTECH, Pohang 37673, Korea}
\date{\small \today}
\providecommand{\keywords}[1]{\textbf{\textit{Keywords   }} #1}

\begin{document}
\maketitle

\begin{abstract}
Using the matrix product ansatz, we obtain solutions of the steady-state distribution of the two-species open one-dimensional zero range process.
Our solution is based on a conventionally employed constraint on the hop rates, which eventually allows us to simplify the constituent matrices of the ansatz.
It is shown that the matrix at each site is given by the tensor product of two sets of matrices and the steady-state distribution assumes an inhomogeneous factorized form.
Our method can be generalized to the cases of more than two species of particles.
\end{abstract}

\maketitle

\section{Introduction}\label{sec:introduction}
Over the last decades, the matrix product ansatz has proven very successful in solving one-dimensional (1D) many-body problems~\cite{Klumper1992ZPB,Derrida1993JPA,Karevski2013PRL,AngelettiBertinAbry2014,Prosen2015JPA}.
In the seminal work~\cite{Derrida1993JPA}, Derrida and coworkers solved the stationary distribution of the open totally asymmetric simple exclusion process using the matrix product ansatz. 
They found very simple algebraic rules of the matrices and obtained an explicit representation of them. 
As a result, they could derive exact expressions for the current and density profiles.
Since then, a lot of developments and generalizations have been made.
For example, a good deal of work has been done for models with multiple species of particles~\cite{Karimipour1999PRE,Uchiyama2008Chaos,Arita2013JPA,Crampe2015JPA,Evans2003JPA,Grosskinsky2003BullBraMath} and the matrix product ansatz has been shown to be closely related to integrable models and tensor-network methods~\cite{Sandow1994PRE,Alcaraz2006JPA,Golinelli2006JPA,Katsura2010JPA,Mei2017PRE,Crampe2018}.

In particular, in~\cite{Krebs1997JPA}, the authors proved that the stationary states of a large class of exclusion processes with open boundaries can be calculated exactly by using the homogeneous matrix product ansatz.
While this proof assumes a finite number of local configurations on each site, recent works obtained the steady-state distributions of different types of generalized zero range processes (ZRP)---stochastic hopping models on a lattice with the hopping rate depending on the occupation number---with periodic boundary conditions~\cite{Chatterjee2017,KunibaOkado2017,KunibaMangazeev2017} and a single-species ZRP with open boundary conditions~\cite{Bertin2018JPA} using the matrix product ansatz with an unbounded number of configurations.  
Given the previous works on models with multiple species of particles~\cite{Karimipour1999PRE,Uchiyama2008Chaos,Arita2013JPA,Crampe2015JPA,Evans2003JPA,Grosskinsky2003BullBraMath,EvansFerrariMallick2009,Vanicat2017J_Stat_Phy,FinnRagoucyVanicat2018}, a naturally arising question is the following: is it possible to obtain the steady-state distribution of the two-species open 1D ZRP using the matrix product method? 
The two-species model is important because it is closely related to many interesting statistical physics phenomena such as the behavior of shaken granular gases in which the grains come in one of two sizes, and the behavior of networks with directed edges (see the review article~\cite{Evans2005JPA} and references therein).

In this paper, we give an affirmative answer to the above question. 
To be specific, we solve the two-species open 1D ZRP under a particular constraint on the hop rates employed in~\cite{Evans2003JPA,Grosskinsky2003BullBraMath,Evans2005JPA}.
Our matrix product solution has a convenient property that the matrix at each site is given by the tensor product of two sets of matrices and the steady-state distribution is given as an inhomogeneous factorized form.
Moreover, our method can be used to deal with open 1D ZRP with more than two species of particles.

\section{Two-species open zero range process}\label{sec:ZRP}


\begin{figure}
\includegraphics[width=\textwidth]{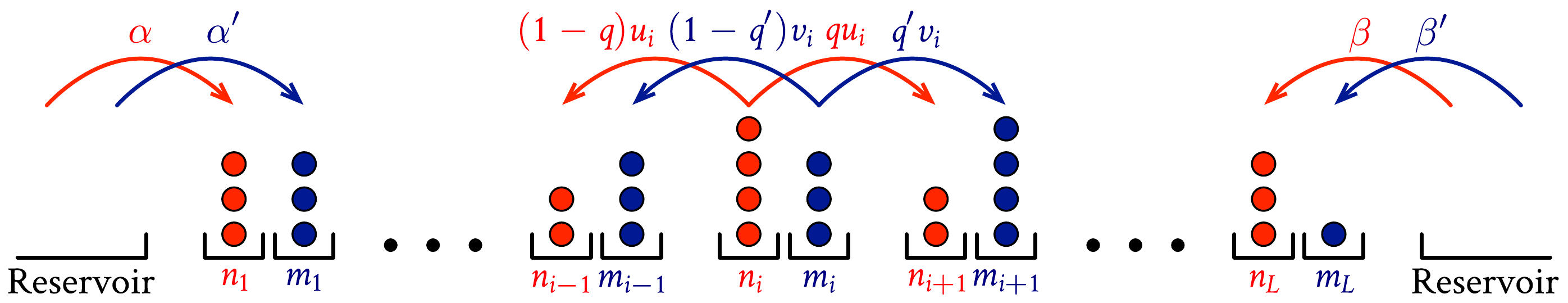}
\caption{(Color online) \small Two-species open 1D zero range process. For brevity, $u(n_i,m_i)$ and $v(n_i,m_i)$ are abbreviated to $u_i$ and $v_i$, respectively.}
\label{fig:zrp}
\end{figure}

We first define the two-species open 1D ZRP. 
Consider a 1D lattice of length $L$, as shown in Figure~\ref{fig:zrp}.
At each site $i$, there are $n_i$ particles of species A and $m_i$ particles of species B. 
The particles are subjected to transport according to the following rule.
One species A particle is transferred from site $i$ to site $i+1$ with rate $qu(n_i,m_i)$ and from site $i$ to site $i-1$ with rate $(1-q)u(n_i,m_i)$, where $0<q<1$ is a parameter of the model and $u(n,m)$ is a function characterizing the rate for species A particles to hop.
Likewise, one species B particle is transferred with rates $q^\prime v(n_i,m_i)$ and $(1-q^\prime )v(n_i,m_i)$, respectively.
At the boundaries, particles are injected from reservoirs to the lattice as follows.
The `left' reservoir injects a species A (B) particle to site $i=1$ with rate $\alpha$ ($\alpha^\prime $) and the `right' reservoir injects to site $i=L$ with rate $\beta$ ($\beta^\prime $).
At the same time, particles at sites $i=1$ and $i=L$ are withdrawn to the reservoirs following the aforementioned rules, i.e., with the rate $(1-q)u(n_1,m_1)$ for species A particles at site $i=1$, and so on.
We set $u(0,0)=v(0,0)=0$.

The master equation governing the probability distribution $P(n_1,m_1,n_2,m_2,...,n_L,\\m_L,t)=P(C,t)$ of the two-species open 1D ZRP is given by~\cite{Blythe2007JPA}
\begin{equation} \label{eq:mastereq_abstract}
\frac{\partial}{\partial t}P(C,t)=\sum_{C^{\prime}\neq C}P(C^{\prime},t)\omega\left(C^{\prime}\rightarrow C\right)-\sum_{C^{\prime}\neq C}P(C,t)\omega\left(C\rightarrow C^{\prime}\right),
\end{equation}
where $C$ and $C^{\prime}$ are two configurations of the particles that differ by a single particle hop, and $\omega\left(C\rightarrow C^{\prime}\right)$ is the rate at which the hop occurs. 
The probability that a hop takes place in an infinitesimal time interval $dt$ is $\omega\left(C\rightarrow C^{\prime}\right)dt$. 
We give the explicit expression of the master equation in appendix~\ref{appendix_1}.

\section{Solution by the matrix product ansatz}

\subsection{Simplification of the master equation by the matrix product ansatz}

We assume that the stationary solution of the master equation (\ref{eq:mastereq_abstract}) can be written as the matrix product ansatz
\begin{equation}\label{eq:matrix_product_ansatz}
P_{st}(n_{1},m_{1},\ldots,n_{L},m_{L})=\frac{1}{Z}\left\langle W\right|R(n_{1},m_{1})\ldots R(n_{L},m_{L})\left|V\right\rangle,
\end{equation}
where $R(n,m)$ is a matrix-valued function of integer variables $n$ and $m$,
$\left\langle W\right|$ and $\left|V\right\rangle$ are boundary vectors, and $Z$ is the normalization constant.
In general, $R(n_{i},m_{i})R(n_{j},m_{j})\neq R(n_{j},m_{j})R(n_{i},m_{i})$ when $i\neq j$.

A rather tedious, but straightforward, calculation (inserting equation~(\ref{eq:matrix_product_ansatz}) into equation~(\ref{eq:mastereq_explicit})) reveals that the stationary solution is obtained if 
\begin{eqnarray}\label{eq:bulk_original}
&&qu(n_{i}+1,m_{i})R(n_{i}+1,m_{i})R(n_{i+1}-1,m_{i+1})     \nonumber \\
&&\quad +(1-q)u(n_{i+1}+1,m_{i+1})R(n_{i}-1,m_{i})R(n_{i+1}+1,m_{i+1})   \nonumber \\
&&\quad +q^{\prime}v(n_{i},m_{i}+1)R(n_{i},m_{i}+1)R(n_{i+1},m_{i+1}-1)   \nonumber \\
&&\quad +(1-q^{\prime})v(n_{i+1},m_{i+1}+1)R(n_{i},m_{i}-1)R(n_{i+1},m_{i+1}+1)   \nonumber \\
&&\quad -\{ qu(n_{i},m_{i}) + (1-q)u(n_{i+1},m_{i+1}) \} R(n_{i},m_{i})R(n_{i+1},m_{i+1})   \nonumber \\
&&\quad -\{ q^{\prime}v(n_{i},m_{i}) + (1-q^{\prime})v(n_{i+1},m_{i+1}) \} R(n_{i},m_{i})R(n_{i+1},m_{i+1})    \nonumber \\
&& = R(n_{i},m_{i})\overline{R}(n_{i+1},m_{i+1})-\overline{R}(n_{i},m_{i})R(n_{i+1},m_{i+1})
\end{eqnarray}
is satisfied along with the boundary conditions
\begin{eqnarray}\label{eq:left_original}
&&\left\langle W\right|[\alpha R(n_{1}-1,m_{1})+(1-q)u(n_{1}+1,m_{1})R(n_{1}+1,m_{1})        \nonumber \\
&&\quad +\alpha^{\prime}R(n_{1},m_{1}-1)+(1-q^{\prime})v(n_{1},m_{1}+1)R(n_{1},m_{1}+1)       \nonumber \\
&&\quad -\{ \alpha + \alpha^{\prime} + (1-q)u(n_{1},m_{1}) + (1-q^{\prime})v(n_{1},m_{1}) \} R(n_{1},m_{1}) ]     \nonumber \\
&& = \left\langle W\right|\overline{R}(n_{1},m_{1})
\end{eqnarray}
and
\begin{eqnarray}\label{eq:right_original}
&&[\beta R(n_{L}-1,m_{L}) + qu(n_{L}+1,m_{L})R(n_{L}+1,m_{L})      \nonumber \\
&&\quad +\beta^{\prime}R(n_{L},m_{L}-1) + q^{\prime}v(n_{L},m_{L}+1)R(n_{L},m_{L}+1)     \nonumber \\
&&\quad - \{ \beta + \beta^\prime + qu(n_{L},m_{L}) + q^{\prime}v(n_{L},m_{L}) \} R(n_{L},m_{L})]\left|V\right\rangle \nonumber\\
&& = -\overline{R}(n_{L},m_{L})\left|V\right\rangle, 
\end{eqnarray}
where $\overline{R}(n,m)$ is an auxiliary matrix, which will be determined later.
Equations (\ref{eq:bulk_original})-(\ref{eq:right_original}) are sufficient conditions for equation $\frac{\partial}{\partial t}P(C,t) = 0$ to be satisfied. In equation (\ref{eq:bulk_original}), $i=1,...,L-1$.

As a next step, we need to simplify equations (\ref{eq:bulk_original})-(\ref{eq:right_original}). 
Motivated by~\cite{Evans2003JPA,Grosskinsky2003BullBraMath,Bertin2018JPA,Evans2005JPA,Levine2005JStatPhy}, we assume that the matrices $R(n,m)$ and $\overline{R}(n,m)$ take the following forms:
\begin{eqnarray}
R(n,m) &=& f(n,m)K(n,m), \label{eq:ansatz_R}   \\
\overline{R}(n,m) &=& f(n,m)\overline{K}(n,m), \label{eq:ansatz_R_bar}
\end{eqnarray}
where
\begin{equation}\label{eq:f_PBC}
f(n,m)=\left\{ \prod_{i=1}^{n}\frac{1}{u(i,m)}\right\} \left\{ \prod_{j=1}^{m}\frac{1}{v(0,j)}\right\}
\end{equation}
and the hop rates satisfy the constraint
\begin{equation}\label{eq:constraint}
\frac{u(n_{i},m_{i})}{u(n_{i},m_{i}-1)}=\frac{v(n_{i},m_{i})}{v(n_{i}-1,m_{i})}.
\end{equation}
Equations (\ref{eq:f_PBC}) and (\ref{eq:constraint}) are sufficient conditions for our ansatz (equations (\ref{eq:ansatz_R}) and (\ref{eq:ansatz_R_bar})) to work.
We note that equation (\ref{eq:constraint}) was obtained in~\cite{Evans2003JPA} for two-species ZRP with the {\em periodic} boundary conditions.
Therein, $f(n,m)$ in equation (\ref{eq:f_PBC}) is closely related to the steady state probability distribution (see equation ($2$) in~\cite{Evans2003JPA}).

Based on equations (\ref{eq:ansatz_R})-(\ref{eq:constraint}), we can reformulate equations (\ref{eq:bulk_original})-(\ref{eq:right_original}) as follows:
\begin{eqnarray} \label{eq:bulk2}
&&u(n_{i+1},m_{i+1}) \{ qK(n_{i}+1,m_{i})K(n_{i+1}-1,m_{i+1})       \nonumber \\
&&\quad\quad\quad\quad\quad\quad-(1-q)K(n_{i},m_{i})K(n_{i+1},m_{i+1}) \}       \nonumber \\
&&\quad +v(n_{i+1},m_{i+1}) \{ q^{\prime}K(n_{i},m_{i}+1)K(n_{i+1},m_{i+1}-1)    \nonumber \\
&&\quad\quad\quad\quad\quad\quad -(1-q^{\prime})K(n_{i},m_{i})K(n_{i+1},m_{i+1}) \}    \nonumber \\
&&\quad +u(n_{i},m_{i}) \{ (1-q)K(n_{i}-1,m_{i})K(n_{i+1}+1,m_{i+1})  \nonumber \\
&&\quad\quad\quad\quad\quad\quad-qK(n_{i},m_{i})K(n_{i+1},m_{i+1}) \}   \nonumber \\
&&\quad +v(n_{i},m_{i}) \{ (1-q^{\prime})K(n_{i},m_{i}-1)K(n_{i+1},m_{i+1}+1)   \nonumber \\
&&\quad\quad\quad\quad\quad\quad-q^{\prime}K(n_{i},m_{i})K(n_{i+1},m_{i+1}) \}    \nonumber \\
&& = K(n_{i},m_{i})\overline{K}(n_{i+1},m_{i+1})-\overline{K}(n_{i},m_{i})K(n_{i+1},m_{i+1}),
\end{eqnarray}
\begin{eqnarray} \label{eq:left2}
&&\langle W|[ u(n_{1},m_{1}) \{ \alpha K(n_{1}-1,m_{1})-(1-q) K(n_{1},m_{1}) \}   \nonumber \\
&&\quad\quad +v(n_{1},m_{1}) \{ \alpha^{\prime} K(n_{1},m_{1}-1)-(1-q^{\prime})K(n_{1},m_{1}) \}   \nonumber \\
&&\quad\quad +(1-q) K(n_{1}+1,m_{1})+(1-q^{\prime}) K(n_{1},m_{1}+1)    \nonumber \\
&&\quad\quad -(\alpha + \alpha^\prime ) K(n_{1},m_{1}) ] \nonumber \\
&& = \left\langle W\right|\overline{K}(n_{1},m_{1}),
\end{eqnarray}
\begin{eqnarray} \label{eq:right2}
&&[u(n_{L},m_{L}) \{ \beta K(n_{L}-1,m_{L}) -qK(n_{L},m_{L}) \}    \nonumber \\
&&\quad +v(n_{L},m_{L}) \{ \beta^{\prime}K(n_{L},m_{L}-1) -q^{\prime}K(n_{L},m_{L}) \}    \nonumber \\
&&\quad +qK(n_{L}+1,m_{L}) +q^{\prime}K(n_{L},m_{L}+1) \nonumber\\
&&\quad -(\beta + \beta^\prime ) K(n_{L},m_{L}) ] \left|V\right\rangle  \nonumber\\
&& = -\overline{K}(n_{L},m_{L})\left|V\right\rangle.  
\end{eqnarray}
In what follows, we further simplify equations (\ref{eq:bulk2})-(\ref{eq:right2}).

\subsection{Parametrization of $K(n,m)$}

Motivated by~\cite{Bertin2018JPA,Mei2017PRE}, we take
\begin{equation}\label{most_important_equation}
K(n,m)=(C_{a}B_{a}^{n})\otimes (D_{b}A_{b}^{m}),
\end{equation}
where $C_{a}$ and $B_{a}$ are matrices living in space $V_a$, while $D_{b}$ and $A_{b}$ are matrices living in another space $V_b$. 
In general, $C_{a}B_{a}\neq B_{a}C_{a}$ and $D_{b}A_{b}\neq A_{b}D_{b}$. 
We will see below that this choice of $K(n,m)$ makes the solutions of equations (\ref{eq:bulk2})-(\ref{eq:right2}) very simple.

Using equation (\ref{most_important_equation}), we can see that the bulk equation (\ref{eq:bulk2}) is satisfied if 
\begin{eqnarray}
qB_{a}C_{a}-(1-q)C_{a}B_{a} &=& \gamma C_{a},   \\
q^{\prime}A_{b}D_{b}-(1-q^{\prime})D_{b}A_{b} &=& \delta D_{b},
\end{eqnarray}
where $\gamma$ and $\delta$ are two arbitrary real numbers and $\overline{K}(n,m)$ is given by
\begin{eqnarray}\label{eq:Kbar}
\overline{K}(n,m) =&\gamma u(n,m) (C_{a}B_{a}^{n-1})\otimes (D_{b}A_{b}^{m})
+\delta v(n,m) (C_{a}B_{a}^{n})\otimes (D_{b}A_{b}^{m-1})  \nonumber\\
&+\lambda (C_{a}B_{a}^{n})\otimes (D_{b}A_{b}^{m}),
\end{eqnarray}
where $\lambda$ is another real number. Using these relations in equation (\ref{eq:left2}), we obtain
\begin{eqnarray}
\left\langle W\right|B_{a} &=& \frac{\alpha}{q}\left\langle W\right|,    \label{eq:leftB}  \\
\left\langle W\right|A_{b} &=& \frac{\alpha^{\prime}}{q^{\prime}}\left\langle W\right|,   \label{eq:leftA}   \\
\lambda &=& -(\gamma+\delta). \label{eq:lambda}
\end{eqnarray}
Using equations (\ref{most_important_equation})-(\ref{eq:Kbar}) in equation (\ref{eq:right2}), we obtain
\begin{eqnarray}
B_{a}\left|V\right\rangle &=& \left(\frac{\beta+\gamma}{q}\right)\left|V\right\rangle , \label{eq:B_right} \\
A_{b}\left|V\right\rangle &=& \left(\frac{\beta^{\prime}+\delta}{q^{\prime}}\right)\left|V\right\rangle,  \label{eq:A_right}
\end{eqnarray}
and equation (\ref{eq:lambda}). 
At this point, $\gamma$ and $\delta$ are arbitrary, but they will be fixed by the boundary conditions later.

\subsection{Results}

Based on equations (\ref{eq:leftB})-(\ref{eq:leftA}) and equations (\ref{eq:B_right})-(\ref{eq:A_right}), we let
\begin{eqnarray}
\left\langle W\right| &=& _{a}\!\!\left\langle W\right|\otimes \, _{b}\!\!\left\langle W\right|,  \\
\left|V\right\rangle &=& \left|V\right\rangle _{a}\otimes\left|V\right\rangle _{b}.
\end{eqnarray}
Using these, we obtain
\begin{eqnarray}
qB_{a}C_{a}-(1-q)C_{a}B_{a} &=& \gamma C_{a},  \nonumber \\
_{a}\!\!\left\langle W\right|B_{a} &=& \left(\frac{\alpha}{q}\right) \,  _{a}\!\!\left\langle W\right|,   \nonumber \\ 
B_{a}\left|V\right\rangle _{a} &=& \left(\frac{\beta+\gamma}{q}\right)\left|V\right\rangle _{a},   \label{eq:BCWV}
\end{eqnarray}
and
\begin{eqnarray}
q^{\prime}A_{b}D_{b}-(1-q^{\prime})D_{b}A_{b} &=& \delta D_{b},    \nonumber \\
_{b}\!\!\left\langle W\right|A_{b} &=& \left(\frac{\alpha^{\prime}}{q^{\prime}}\right) \, _{b}\!\!\left\langle W\right|,     \nonumber \\
A_{b}\left|V\right\rangle _{b} &=& \left(\frac{\beta^{\prime}+\delta}{q^{\prime}}\right)\left|V\right\rangle _{b}. \label{eq:ADWV}
\end{eqnarray}
Then, equation (\ref{eq:matrix_product_ansatz}) becomes
\begin{eqnarray}
&&P_{st}(n_{1},m_{1},...,n_{L},m_{L}) \nonumber \\
&&=\frac{1}{Z}\left\{\prod_{j=1}^{L}f(n_{j},m_{j})\right\}    \nonumber \\
&&\quad \times \left({}_{a}\!\left\langle W\right|C_{a}B_{a}^{n_{1}}\ldots C_{a}B_{a}^{n_{L}}\left|V\right\rangle _{a}\right)   \left( {}_{b}\!\left\langle W\right|D_{b}A_{b}^{m_{1}}\ldots D_{b}A_{b}^{m_{L}}\left|V\right\rangle _{b}\right).
\end{eqnarray}
Note that the stationary distribution does not factorize into two distributions associated with independent single-species models because the two species of particles are strongly coupled through $f(n_j,m_j)$.

At this stage, we can use the results in~\cite{Bertin2018JPA} to finish the calculation. 
The dimension of space $V_a$ is $L+1$ and the basis vectors are given by $\left|k\right\rangle _{a}$ ($k=0,...,L$). 
We have
\begin{eqnarray}
_{a}\!\bra{k}\ket{k^{\prime}}_{a} &=&\delta_{k k^{\prime}},   \nonumber \\
C_{a} &=&\sum_{k=1}^{L}\left|k-1\right\rangle _{a}  \cdot  \, _{a}\!\left\langle k\right| = \left(\begin{array}{ccccc}
0 & 1 & 0 & \cdots & 0\\
\vdots & \ddots & \ddots & \ddots & \vdots\\
\vdots &  & \ddots & \ddots & 0\\
\vdots &  &  & \ddots & 1\\
0 & \cdots & \cdots & \cdots & 0
\end{array}\right)_{a},  \nonumber \\
B_{a} &=&\sum_{k=0}^{L}x_{k}\left|k\right\rangle _{a}  \cdot  \,  _{a}\!\left\langle k\right|=\left(\begin{array}{cccc}
x_{0} & 0 & \cdots & 0\\
0 & x_{1} & \ddots & \vdots\\
\vdots & \ddots & \ddots & 0\\
0 & \cdots & 0 & x_{L}
\end{array}\right)_{a},  \nonumber \\
_{a}\!\left\langle W\right| &=& \, _{a}\!\left\langle 0\right|=\left(\begin{array}{cccc}
1 & 0 & \cdots & 0\end{array}\right)_{a},   \nonumber \\  
\left|V\right\rangle _{a} &=& \left|L\right\rangle _{a}=\left(\begin{array}{c}
0\\
\vdots\\
0\\
1
\end{array}\right)_{a}.  \label{representation_a}
\end{eqnarray}
Similar results for space $V_b$ can be obtained by making changes $a \rightarrow b$, $C \rightarrow D$, $B \rightarrow A$, and $x \rightarrow y$.
Using equation (\ref{representation_a}) in equation (\ref{eq:BCWV}), we obtain
\begin{eqnarray}
q\cdot x_{k-1}-(1-q)\cdot x_{k} &=& \gamma,     \nonumber \\  
x_{0} &=& \frac{\alpha}{q},        \nonumber \\   
x_{L} &=& \frac{\beta+\gamma}{q},
\end{eqnarray}
where $k=1,...,L$. Solving the above equations, we obtain
\begin{eqnarray}
\gamma &=& (2q-1)\frac{\alpha q^{L}-\beta(1-q)^{L}}{q^{L+1}-(1-q)^{L+1}},   \\
x_{m} &=& \frac{\alpha}{q}\left(\frac{q}{1-q}\right)^{m}+\left[1-\left(\frac{q}{1-q}\right)^{m}\right]\frac{\alpha q^{L}-\beta(1-q)^{L}}{q^{L+1}-(1-q)^{L+1}}.
\end{eqnarray}
Similarly, we obtain from equation (\ref{eq:ADWV}) that
\begin{eqnarray}
\delta &=& (2q^{\prime}-1)\frac{\alpha^{\prime}(q^{\prime})^{L}-\beta^{\prime}(1-q^{\prime})^{L}}{(q^{\prime})^{L+1}-(1-q^{\prime})^{L+1}},   \\
y_{m} &=& \frac{\alpha^{\prime}}{q^{\prime}}\left(\frac{q^{\prime}}{1-q^{\prime}}\right)^{m}+\left[1-\left(\frac{q^{\prime}}{1-q^{\prime}}\right)^{m}\right]\frac{\alpha^{\prime}(q^{\prime})^{L}-\beta^{\prime}(1-q^{\prime})^{L}}{(q^{\prime})^{L+1}-(1-q^{\prime})^{L+1}}.
\end{eqnarray}
We remark that the parameters $x_m$ and $y_m$ appeared in the solution of the open single-species ZRP in~\cite{Levine2005JStatPhy}.

After straightforward calculations, the steady-state distribution $P_{st}(n_{1},m_{1},\ldots,n_{L},m_{L})$ of the two-species open 1D ZRP is finally given by the following inhomogeneous factorized form: 
\begin{equation}
P_{st}(n_{1},m_{1},\ldots,n_{L},m_{L})=\prod_{k=1}^{L}P_{k}(n_{k},m_{k})
\end{equation}
with
\begin{equation}
P_{k}(n,m)=f(n,m)\frac{(x_{k})^{n}(y_{k})^{m}}{Z_{k}},
\end{equation}
where $Z_{k}$ is a normalization factor determined by 
\begin{equation}
\sum_{n=0}^{+\infty}\sum_{m=0}^{+\infty}P_{k}(n,m)=1.
\end{equation}

The key step in the derivation was to introduce equation (\ref{most_important_equation}). 
One may choose a seemingly different matrix ansatz
\begin{equation}\label{another_ansatz}
K(n,m)=BA_{1}^{n}A_{2}^{m},
\end{equation}
where $BA_{1}\neq A_{1}B$, $BA_{2}\neq A_{2}B$, and $A_{1}A_{2}=A_{2}A_{1}$. 
However, one can easily build the connection between equations (\ref{most_important_equation}) and (\ref{another_ansatz}). 
Rewriting equation (\ref{most_important_equation}) as
\begin{equation}
K(n,m)=\left(C_{a}\otimes D_{b}\right)\cdot\left(B_{a}\otimes I_{b}\right)^{n}\cdot\left(I_{a}\otimes A_{b}\right)^{m}
\end{equation}
with $I_{a}$ ($I_{b}$) being the identity matrix in space $V_a$ ($V_b$), we can see that $B=C_{a}\otimes D_{b}$, $A_1=B_{a}\otimes I_{b}$, and $A_2=I_{a}\otimes A_{b}$.

\section{Case study of the model}

At site $k$, the probability to find $n_k$ $A$ particles and $m_k$ $B$ particles is $P_{k}(n_k,m_k)$. The average number of $A$ particles at site $k$ is
\begin{equation}\label{eq:N_A_k_1}
\left\langle N_{A,k}\right\rangle =\sum_{n=0}^{+\infty}\sum_{m=0}^{+\infty}n\cdot P_{k}(n,m).
\end{equation}
Similarly, the average number of $B$ particles at site $k$ is
\begin{equation}
\left\langle N_{B,k}\right\rangle =\sum_{n=0}^{+\infty}\sum_{m=0}^{+\infty}m\cdot P_{k}(n,m).
\end{equation}
For simplicity, we take~\cite{Evans2005JPA}
\begin{equation}\label{v_n_m}
v(n,m)=1+\frac{c}{(n+1)^{\eta}},
\end{equation}
where $c$, $\eta >0$. Choosing $u(n,0)=1$, we obtain from equation (\ref{eq:constraint})
\begin{equation}
u(n,m)=\left[\frac{1+\frac{c}{(n+1)^{\eta}}}{1+\frac{c}{n^{\eta}}}\right]^{m},
\end{equation}
leading to
\begin{equation}\label{eq:f_n_m_final}
f(n,m)=\left[1+\frac{c}{(n+1)^{\eta}}\right]^{-m}.
\end{equation}
From equations (\ref{eq:N_A_k_1})-(\ref{eq:f_n_m_final}), we obtain
\begin{eqnarray}
Z_{k} &=& \sum_{n=0}^{+\infty}\left(x_{k}\right)^{n} \cdot \frac{(n+1)^{\eta}+c}{\left(1-y_{k}\right)\cdot(n+1)^{\eta}+c}, \label{Z_k_final} \\
\left\langle N_{A,k}\right\rangle &=& \frac{1}{Z_{k}}\sum_{n=0}^{+\infty}n\left(x_{k}\right)^{n} \cdot \frac{(n+1)^{\eta}+c}{\left(1-y_{k}\right)\cdot(n+1)^{\eta}+c}, \label{N_A_k_final}\\
\left\langle N_{B,k}\right\rangle &=& \frac{1}{Z_{k}}\sum_{n=0}^{+\infty}\left(x_{k}\right)^{n}\cdot\frac{y_{k}\left[(n+1)^{\eta}+c\right](n+1)^{\eta}}{\left[\left(1-y_{k}\right)\cdot(n+1)^{\eta}+c\right]^{2}}.  \label{N_B_k_final}
\end{eqnarray}
For the series to be convergent, we should have $0<x_k<1$ and $0<y_k<1$. 

In the case of the periodic boundary conditions, as the particle numbers of $A$ and $B$ are conserved, the particle densities are usually used as parameters to investigate the dynamics~\cite{Evans2005JPA}.
In the case of the open 1D ZRP, on the other hand, the particle numbers are not conserved.
In this case, it is more interesting to choose the strength of the boundary drive as a parameter.

\begin{figure}[t]
\centering
\begin{subfigure}{0.5\textwidth}
\centering
\includegraphics[width=0.8\textwidth]{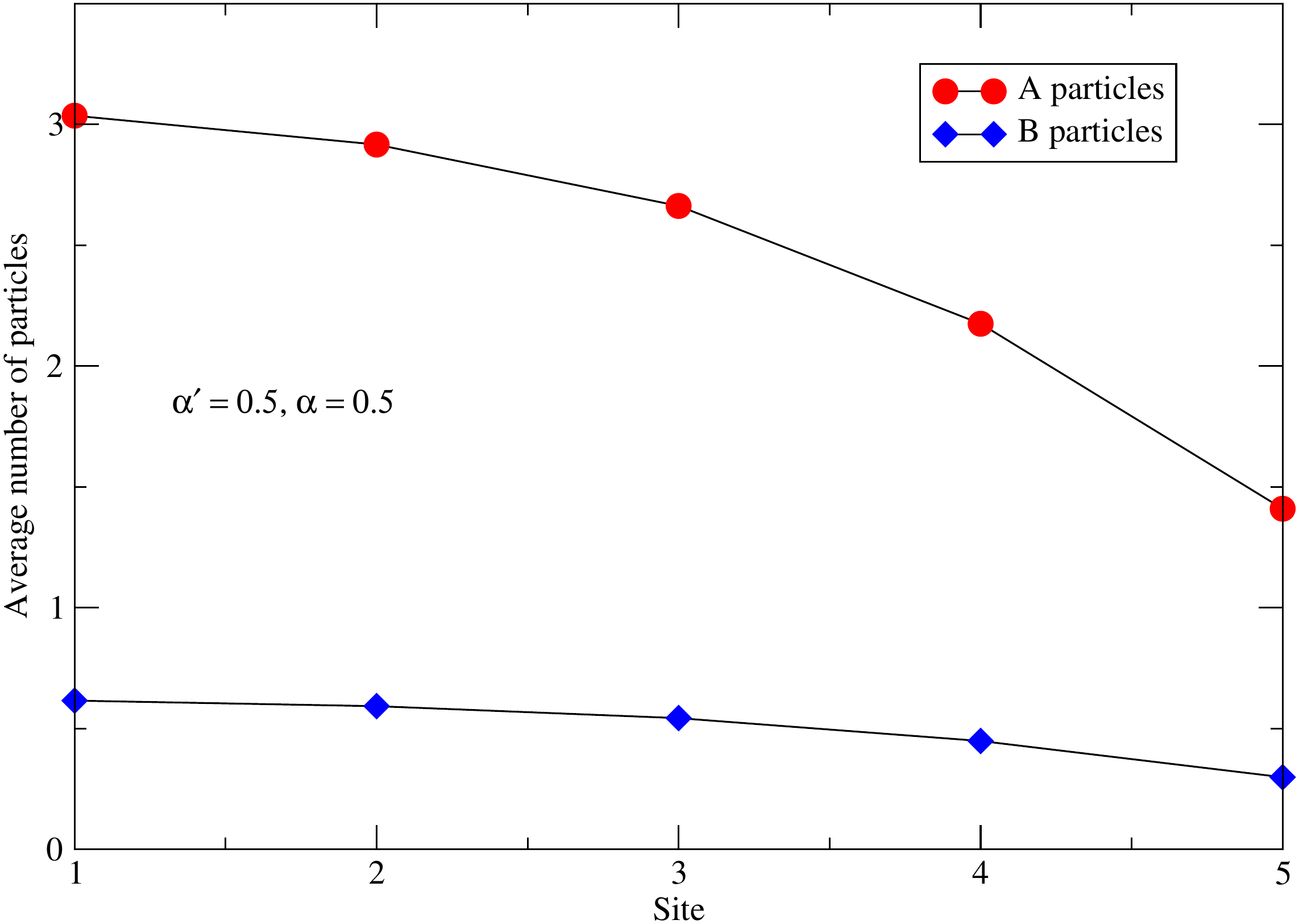}
\caption{\small $\alpha^{\prime}=0.5$, $\alpha=0.5$.}
\label{fig1a}
\end{subfigure}%
\begin{subfigure}{0.5\textwidth}
\centering
\includegraphics[width=0.8\textwidth]{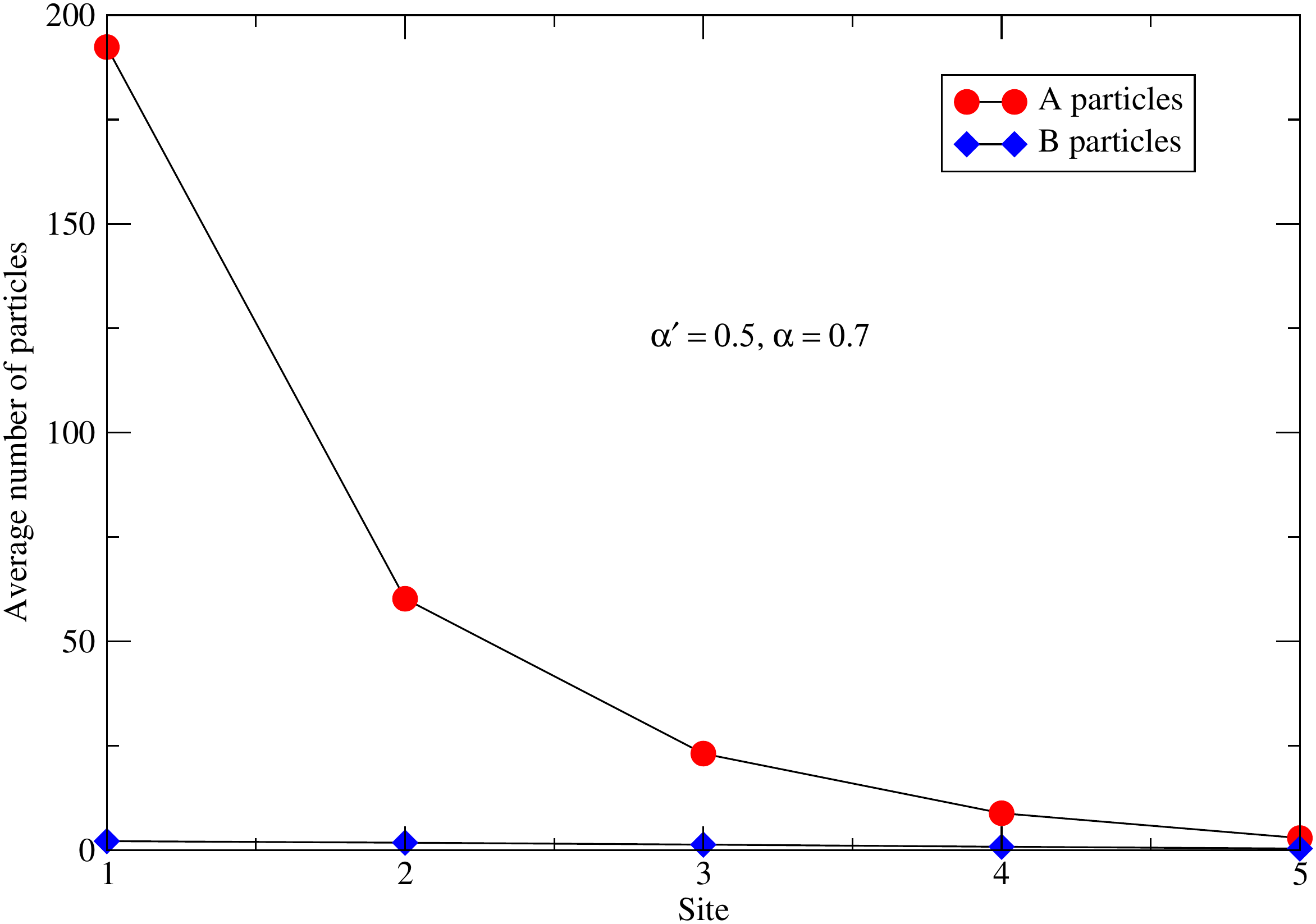}
\caption{\small $\alpha^{\prime}=0.5$, $\alpha=0.7$.}
\label{fig1b}
\end{subfigure} 
\begin{subfigure}{0.5\textwidth}
\centering
\includegraphics[width=0.8\textwidth]{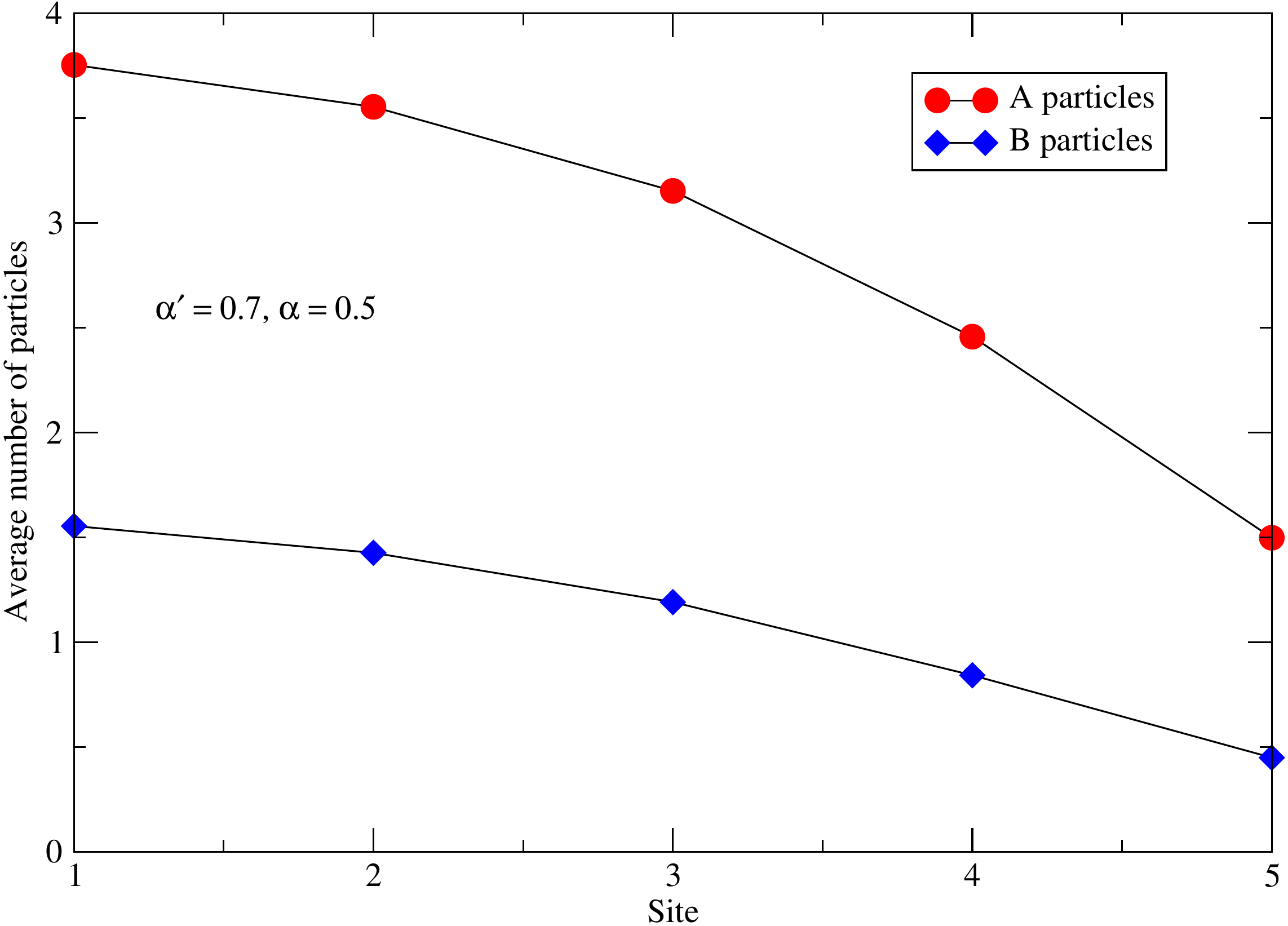}
\caption{\small $\alpha^{\prime}=0.7$, $\alpha=0.5$.}
\label{fig1c}
\end{subfigure}%
\begin{subfigure}{0.5\textwidth}
\centering
\includegraphics[width=0.8\textwidth]{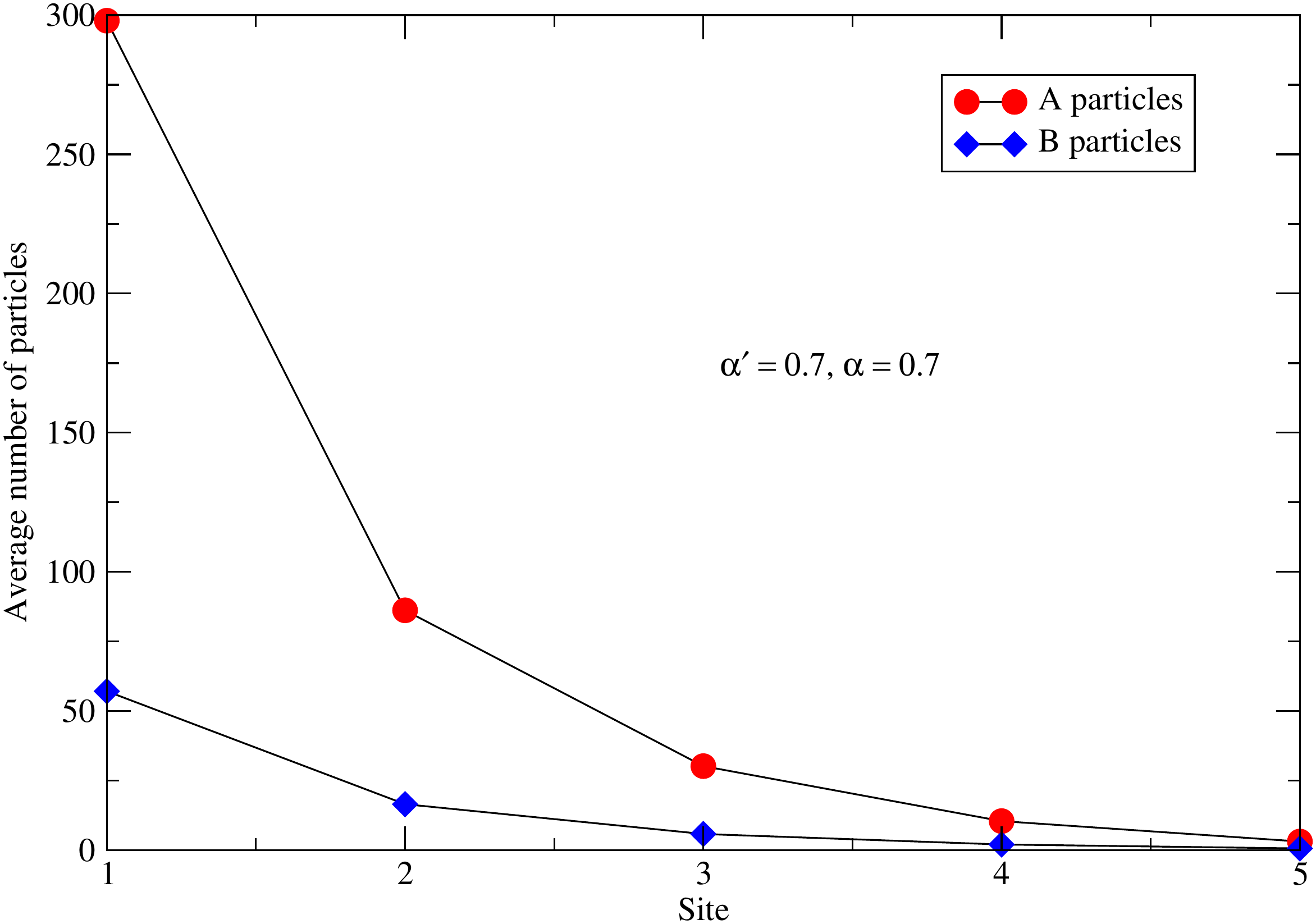}
\caption{\small $\alpha^{\prime}=0.7$, $\alpha=0.7$.}
\label{fig1d}
\end{subfigure}
\caption{(Color online) \small Stationary distribution of particles for the two-species open 1D ZRP. The horizontal axis indicates the site and the vertical axis indicates the average number of particles.
Red circles represent $A$ particles and blue diamonds represent $B$ particles.
Note that the vertical scales are all different.}
\label{fig:results}
\end{figure}

Figure~\ref{fig:results} shows our numerical results, where we fix $c=3$, $\eta=1$, $L=5$, $q=q^{\prime}=0.7$, and $\beta=\beta^{\prime}=0.1$, while $\alpha^{\prime}$ and $\alpha$ are varied.
Recall that $\alpha$ ($\alpha^{\prime}$) denotes the rate for the `left' reservoir to inject one $A$ ($B$) particle to site $1$.
The stationary distributions of particles shown in Figure~\ref{fig:results} reveal nontrivial features of the model.
First, the distribution of one species of particles is strongly influenced by the other due to the interaction between the two species.
For example, the results in (c) and (d) are significantly different although $\alpha^\prime$ is the same and $\alpha$ is only moderately different.
Another interesting phenomenon is that most of the particles are accumulated near site 1 when the system is driven strongly at the boundary, i.e., for large $\alpha$ and $\alpha'$.

\section{Conclusion}

We have derived the steady-state distribution of the two-species open 1D ZRP under the constraint equation (\ref{eq:constraint}). The key step in the derivation is to express the matrix at each site as the tensor product of two sets of matrices. 
The algebraic structure of our solution allows a natural generalization to the cases of more than two species of particles. 
For example, when three species of particles are involved, one should change $R(n,m)$ in equation (\ref{eq:matrix_product_ansatz}) to $R(n,m,l)$ and then generalize the constraint equation (\ref{eq:constraint}) to the one given in~\cite{Grosskinsky2003BullBraMath}.
In addition, one should generalize equation (\ref{most_important_equation}) to $K(n,m,l)=(C_{a}B_{a}^{n})\otimes (D_{b}A_{b}^{m}) \otimes (F_{c}G_{c}^{l})$.
Except for these key steps, all other formulas are similar to those given in this paper.
Based on our analytical results, we have also investigated the effect of boundary drive and found out that the interaction between the two species of particles significantly influence the stationary distribution.

\section*{Acknowledgments}
This research was supported by the R\&D Convergence Program of NST (National Research Council of Science and Technology), the Ministry of Science, ICT \& Future Planning, Gyeongsangbuk-do and Pohang City (Grant No. CAP-15-08-KRISS).

\appendix

\section{Master equation for the two-species open \\one-dimensional zero range process}\label{appendix_1}

According to equation (\ref{eq:mastereq_abstract}), one can write down the master equation for the two-species open 1D ZRP as follows:
\begin{eqnarray}\label{eq:mastereq_explicit}
&&\frac{\partial}{\partial t}P\left(n_{1},m_{1},\ldots,n_{L},m_{L},t\right)  \nonumber\\
&&=\alpha P\left(n_{1}-1,m_{1},\ldots,n_{L},m_{L},t\right)
+\alpha^{\prime}P(n_{1},m_{1}-1,\ldots,n_{L},m_{L},t)    \nonumber \\
&&\quad +\beta P(n_{1},m_{1},\ldots,n_{L}-1,m_{L},t)
+\beta^{\prime}P(n_{1},m_{1},\ldots,n_{L},m_{L}-1,t)           \nonumber \\
&&\quad +(1-q)u(n_{1}+1,m_{1})P(n_{1}+1,m_{1},\ldots,n_{L},m_{L},t)   \nonumber \\
&&\quad +(1-q^{\prime})v(n_{1},m_{1}+1)P(n_{1},m_{1}+1,\ldots,n_{L},m_{L},t) \nonumber \\
&&\quad +qu(n_{L}+1,m_{L})P(n_{1},m_{1},\ldots,n_{L}+1,m_{L},t) \nonumber \\
&&\quad +q^{\prime}v(n_{L},m_{L}+1)P(n_{1},m_{1},\ldots,n_{L},m_{L}+1,t)   \nonumber \\
&&\quad +\sum_{i=1}^{L-1}qu(n_{i}+1,m_{i})P(\ldots,n_{i}+1,m_{i},n_{i+1}-1,m_{i+1},\ldots,t)    \nonumber \\
&&\quad +\sum_{i=1}^{L-1}(1-q)u(n_{i+1}+1,m_{i+1})P(\ldots,n_{i}-1,m_{i},n_{i+1}+1,m_{i+1},\ldots,t)   \nonumber \\
&&\quad +\sum_{i=1}^{L-1}q^{\prime}v(n_{i},m_{i}+1)P(\ldots,n_{i},m_{i}+1,n_{i+1},m_{i+1}-1,\ldots,t)   \nonumber \\
&&\quad +\sum_{i=1}^{L-1}(1-q^{\prime})v(n_{i+1},m_{i+1}+1)P(\ldots,n_{i},m_{i}-1,n_{i+1},m_{i+1}+1,\ldots,t)      \nonumber \\
&&\quad -(\alpha +\alpha^\prime + \beta + \beta^\prime )P(n_{1},m_{1},\ldots,n_{L},m_{L},t)       \nonumber \\
&&\quad -\sum_{i=1}^{L} \{ u(n_{i},m_{i}) + v(n_i, m_i)\}P(n_{1},m_{1},\ldots,n_{L},m_{L},t).
\end{eqnarray}




\end{document}